\documentclass[fleqn,twoside]{article}
\usepackage{espcrc2}

\usepackage{epsfig,multicol,multirow,subfigure}

\usepackage{graphicx}
\usepackage[figuresright]{rotating}

\def\lsi{\raise0.3ex\hbox{$<$\kern-0.75em\raise-1.1ex\hbox{$\sim$}}}
\def\gsi{\raise0.3ex\hbox{$>$\kern-0.75em\raise-1.1ex\hbox{$\sim$}}}

\newcommand{\gsim}{\mathop{\gsi}}

\newcommand{\eps}{\epsilon}
\newcommand{\la}{\langle}
\newcommand{\ra}{\rangle}

\newcommand{\beq}{\begin{equation}}
\newcommand{\eeq}{\end{equation}}

\newcommand{\AmS}{{\protect\the\textfont2
  A\kern-.1667em\lower.5ex\hbox{M}\kern-.125emS}}

\title{Overlap Hypercube Fermions in QCD}

\author{W. Bietenholz \address{ Institut f{\"{u}}r Physik,
Humboldt Universit{\"{a}}t zu Berlin,
Newtonstr.\ 15, D-12489 Berlin, Germany} 
and S. Shcheredin \address{ Fakult{\"{a}}t f{\"{u}}r Physik,
Universit\"{a}t Bielefeld,
D-33615 Bielefeld, Germany}
\thanks{This work was supported by the Deutsche Forschungsgemeinschaft 
through SFB/TR9-03. 
The computations were performed on the IBM p690 clusters of the
``Norddeutscher Verbund f\"ur Hoch- und H\"ochstleistungsrechnen'' (HLRN) 
and at NIC, Forschungszentrum J\"{u}lich.} 
} 

\begin{document}

\begin{abstract}

We present simulation results obtained with overlap hypercube fermions
in QCD near the chiral limit. We relate our results to chiral
perturbation theory in both, the
$\epsilon$-regime and in the $p$-regime. In particular we measured
the pion decay constant by different methods, as well as the chiral 
condensate, light meson masses, the PCAC quark mass and the renormalisation
constant $Z_{A}$.

\vspace{-1pc}
\end{abstract}

\maketitle

\section{INTRODUCTION}

We report on our recent simulation results
for QCD on the lattice with chirally
symmetric fermions. The aim is to relate our numerical results
to chiral perturbation theory ($\chi$PT), as 
an effective theory of the strong interaction at low energies.
The simulation of QCD, as the fundamental theory, allows for a
determination of free parameters in this effective theory.

In Section 2 we review the construction of the hypercube fermion
(HF) and show recent applications at finite temperature. We then
proceed to the corresponding overlap hypercube operator, and we 
present results on its locality at strong gauge coupling.

In Section 3 we turn our interest to the
{\em $\epsilon$-regime}, which is the main focus
of this work.
In this setting, the linear size of the volume is shorter than
the correlation length (which is given by the inverse pion mass). 
As an important virtue, evaluations in this regime do not need an 
extrapolation to the infinite volume: physical quantities 
--- in particular the Low Energy Constants (LEC) of the chiral Lagrangian ---
can be identified directly in a small box, 
with their values in infinite volume.

In Section 4 we add results obtained 
in the (standard) $p$-regime, where the $\chi$PT works
by expanding in the momenta of the light mesons involved,
and in Section 5 we arrive at our conclusions.

Throughout this work we refer to quenched simulations
using the (standard) Wilson gauge action,
and we deal with a physical volume of
$V = (1.48~{\rm fm})^{3} \times 2.96~{\rm fm}$
at two lattice spacings $a$ (measured by the Sommer scale \cite{RSo}).
Preliminary results were presented before in Refs.\ \cite{prelim}.

\section{OVERLAP HYPERCUBE FERMIONS}

For free fermions, {\em perfect lattice actions} are known analytically
\cite{perfect}.
The corresponding lattice Dirac operator can be written as
\beq
D_{x,y} = \gamma_{\mu} \rho_{\mu}(x-y) + \lambda (x-y) \ ,
\eeq
with closed expression for $\rho_{\mu}$ and $\lambda$ in
momentum space. Such actions emerge from iterated renormalisation 
group transformations, hence they are free of any lattice artifacts
regarding scaling and chirality. 
$D_{x,y}$ is local (for non-vanishing $\lambda$) but its range is
infinite. Hence we tuned the renormalisation group parameters
for an optimal locality of $D_{x,y}$; then we truncate it
by means of periodic boundary conditions so that the supports of
$\rho_{\mu}(x-y)$ and $\lambda (x-y)$ are contained in
$\{ | x_{\nu} - y_{\nu}| \leq 1\}$ for $\nu = 1 \dots 4$
(in lattice units, $a=1$).
This truncation yields the free Hypercube Fermion (HF) which still
has excellent scaling and chirality properties \cite{BBCW}.

The HF is gauged by (normalised) sums over the shortest lattice paths 
between its sites. The paths correspond to products of fat link gauge 
variables, which are simply built as $U_{\mu}(x) \to
(1 - \alpha ) U_{\mu}(x) + \frac{\alpha}{6} \sum {\rm staples}$.
Finally the links are amplified to restore criticality
and to minimise the violation of the Ginsparg-Wilson relation.

\begin{figure}
\centering
\includegraphics[angle=0,width=0.9\linewidth]{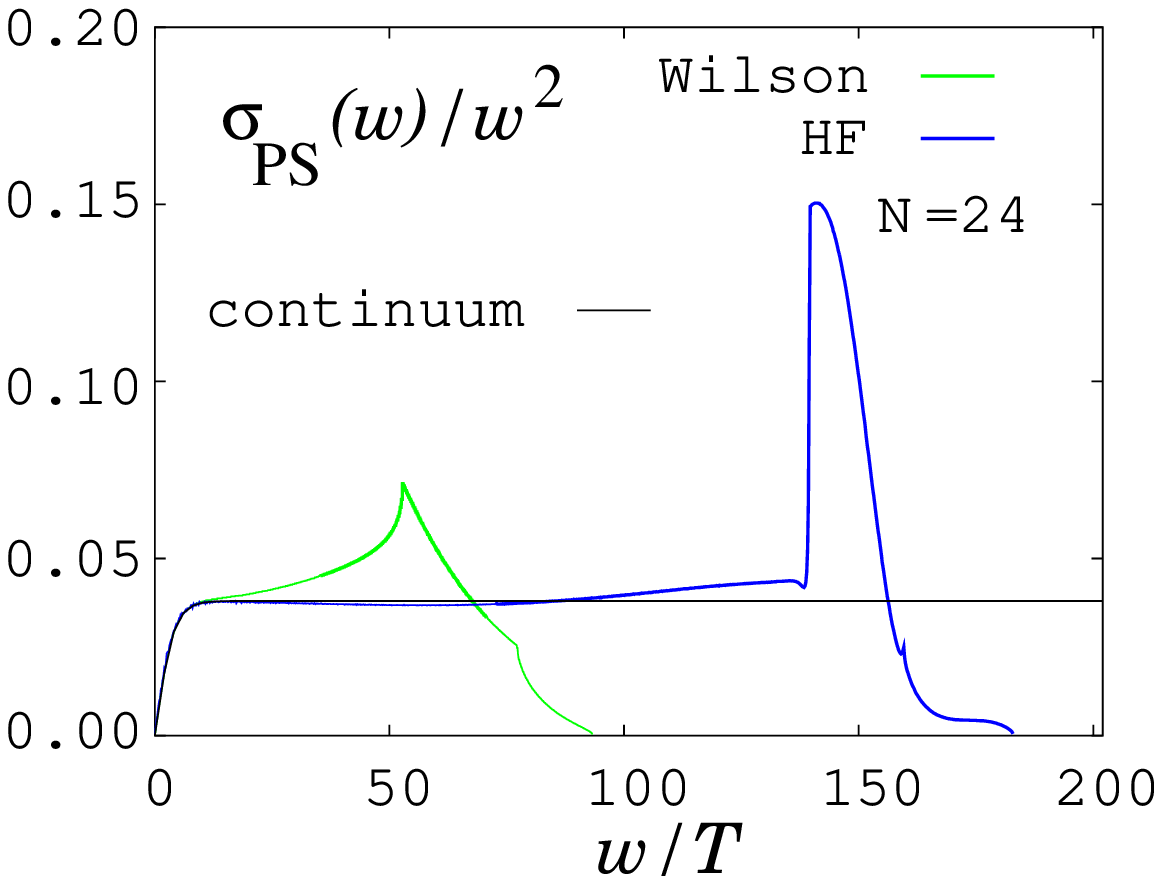} \\
\vspace*{2mm}
\includegraphics[angle=0,width=0.89\linewidth]{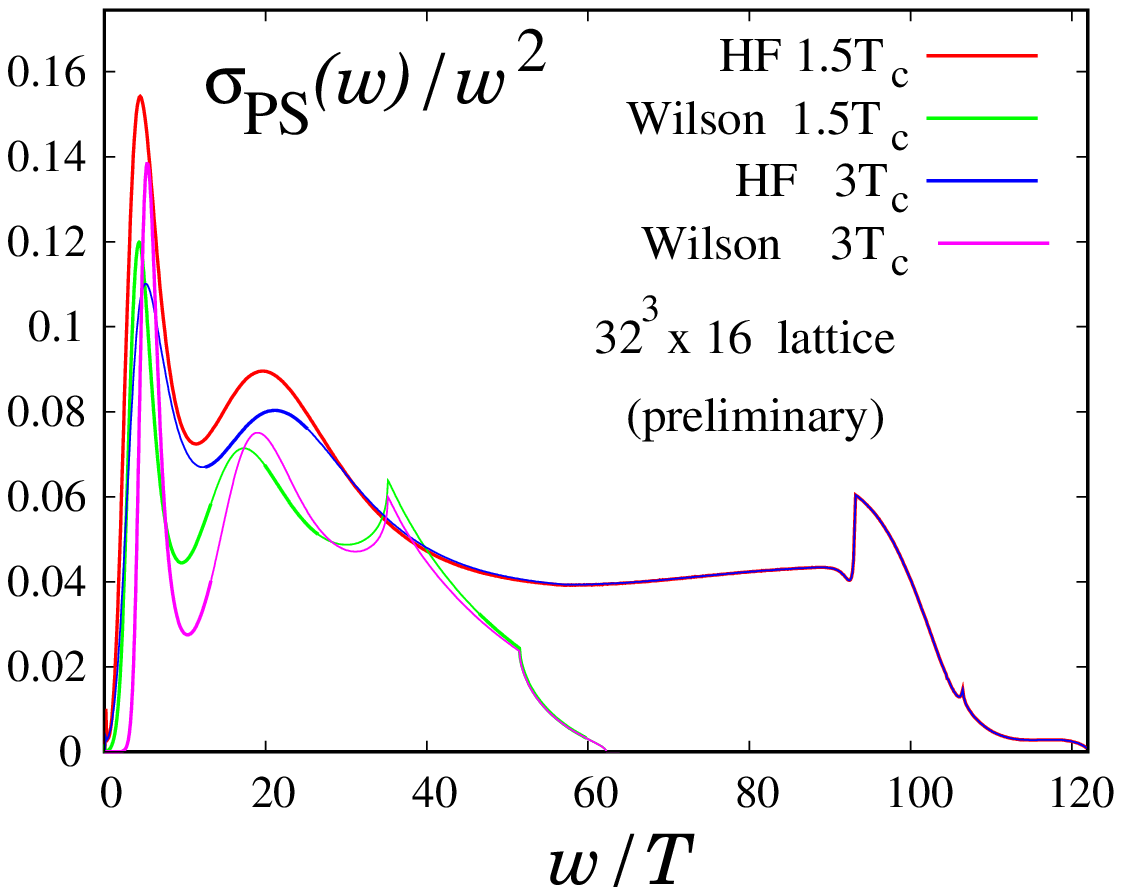}
\vspace*{-7mm}
\caption{{\it The spectral function $\sigma_{\rm PS}$,
as a function of the Matsubara frequencies $\omega$,
at $T_{c} = \infty$ (on top, free case) and at finite $T_{c}$
(below, interacting case) \cite{thermo}. These results are obtained 
from lattice data using the Maximum Entropy Method, as suggested in 
Ref.\ \cite{MEM}.}}
\label{thermofig}
\vspace*{-8mm}
\end{figure}

Due to the truncation and the imperfect gauging procedure, the
scaling and chirality are somewhat distorted.
Still the HF has highly favourable features, particularly
in thermodynamics \cite{BBCW,thermo}.
As an example, we show in Fig.\ \ref{thermofig}
recent results for the spectral function $\sigma_{\rm PS}$,
which reveal a continuum-like behaviour up to much larger energies
than it emerges for the Wilson fermion.

Exact chirality can be recovered by inserting the HF
into the overlap formula \cite{KikNeu}
\beq  \label{overlap}
D_{\rm ov}^{(0)} = \rho (1 + A / \sqrt{A^{\dagger}A} ) \ , \quad
A := D_{0} - \rho \ ,
\eeq
where $\rho \gsim 1 $ and
$D_{0}$ is some lattice Dirac operator, with 
$D_{0} = \gamma_{5} D_{0}^{\dagger}\gamma_{5}$ \ ($\gamma_{5}$-Hermiticity).

$\bullet$ \ The standard formulation \cite{Neu} inserts the
Wilson Dirac operator, $D_{0} = D_{\rm W}$, 
which is then drastically changed in eq.\ (\ref{overlap}).
We denote the resulting {\em standard overlap operator}, or
{\em Neuberger operator}, as $D_{\rm ov-W}$.

$\bullet$ \ Here we mainly study the case where instead the HF is 
inserted in the overlap
formula (\ref{overlap}), $D_{0} = D_{\rm HF}$ \cite{ovHF}. This yields
the operator $D_{\rm ov-HF}$, which describes the {\em overlap-HF}.

In both cases one arrives at exact solutions to the Ginsparg-Wilson relation
\cite{GW}, and therefore at an exact, lattice modified chiral symmetry
\cite{Has-Lusch}.\footnote{The correctness of the axial anomaly in all 
topological sectors has been verified 
for the standard overlap operator in Ref.\ \cite{DHA}, and for the
overlap HF in Ref.\ \cite{AB}.} 
However, in contrast to $D_{\rm W}$, $D_{\rm HF}$ is approximately chiral
already, hence its transformation by the overlap formula (\ref{overlap}),
$D_{\rm HF} \to D_{\rm ov-HF}$, is only a modest modification.
Therefore, the virtues of the HF are essentially inherited by
the overlap-HF. 
Compared to  $D_{\rm ov-W}$,
the degrees of locality and of approximate
rotation symmetry are highly improved \cite{ovHF,WBIH,WB02,Stani}.
If the gauge coupling is not too strong,
the maximal impact of a unit source on a sink, 
as well as the maximal violation of rotation 
symmetry, decay exponentially in the distance.
For instance, at $\beta =6$ the exponent of these decays
is almost doubled for the overlap-HF compared to $D_{\rm ov-W}$ \cite{WB02}.
(In the latter case, locality was first studied in Ref.\ \cite{HJL}).

\begin{figure}[hbt]
\centering
\includegraphics[angle=270,width=0.9\linewidth]{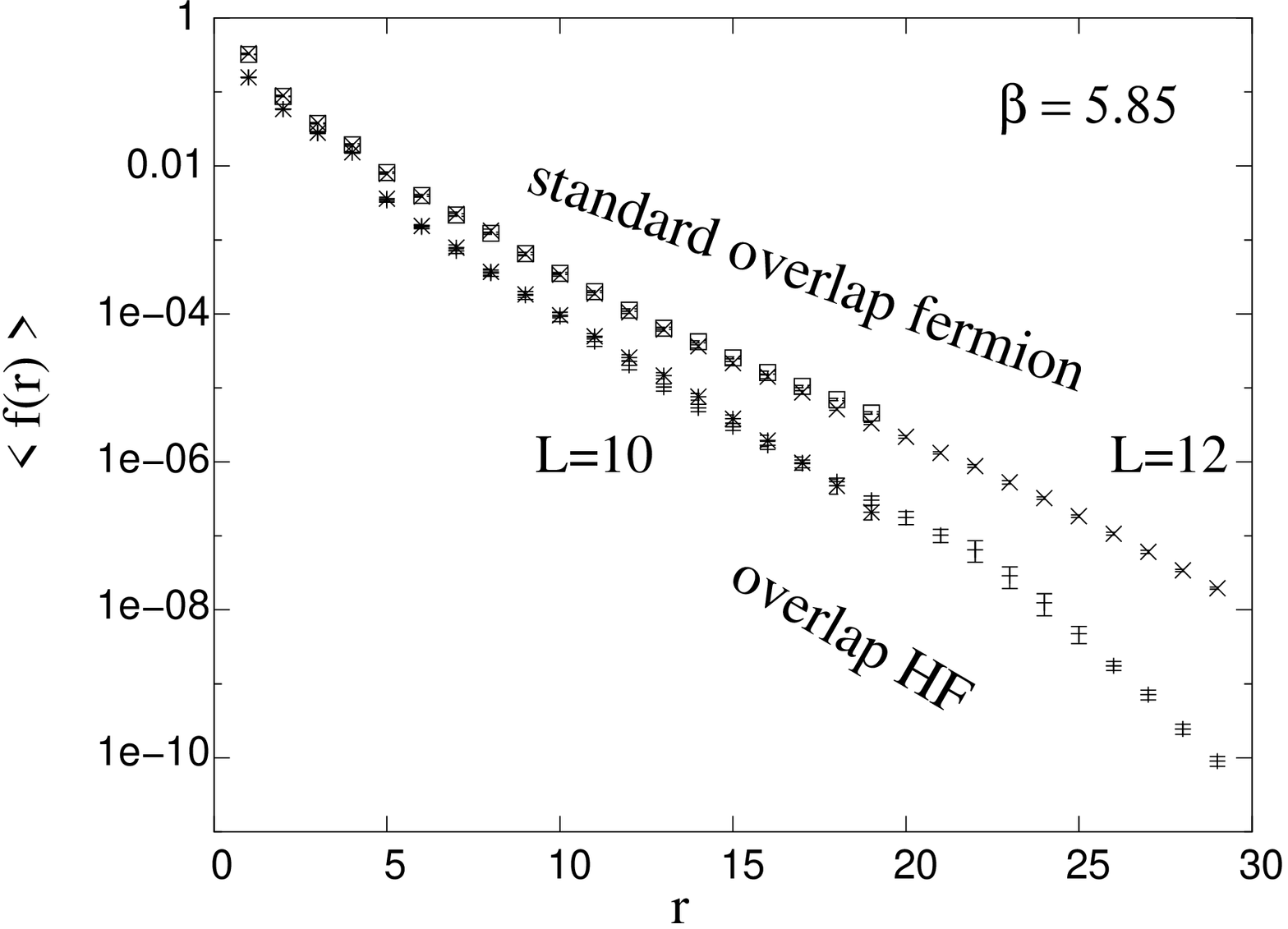}
\includegraphics[angle=270,width=1.\linewidth]{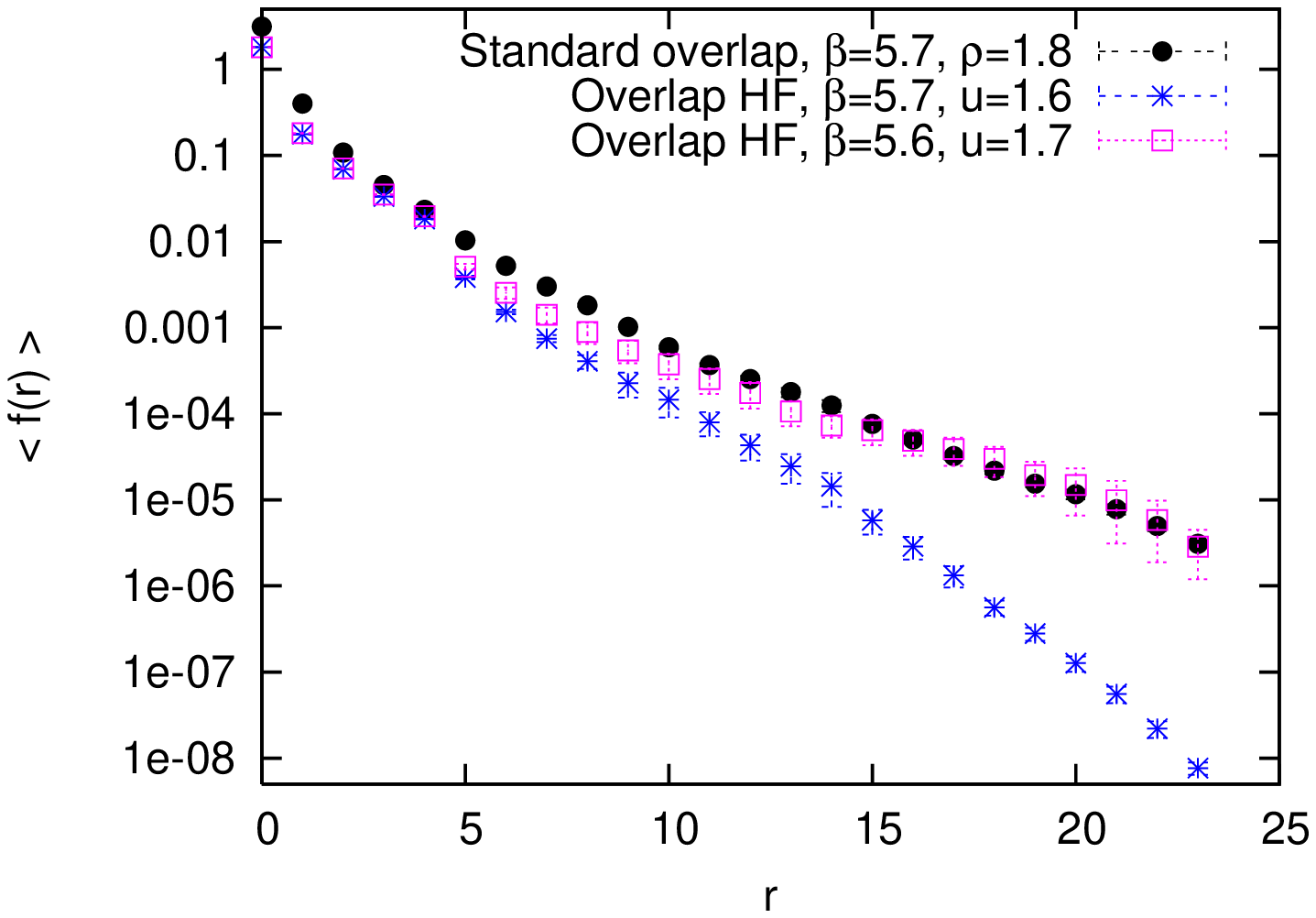}
\vspace*{-8mm}
\caption{{\it The locality of different overlap fermions, measured by the 
maximal impact of a unit source $\bar \psi_{x}$ on $\psi_{y}$ at a distance
$r = \Vert x -y \Vert_{1}$. At the same value of $\beta$, the overlap-HF
is clearly more local, as the plot on top shows for $\beta = 5.85$.
Below we see that its locality persist up to $\beta =5.6$,
where no exponential decay is visible for the standard overlap formulation.}}
\label{localfig}
\vspace*{-10mm}
\end{figure}

Here we are going to show mostly results at $\beta = 5.85$, which corresponds 
to a lattice spacing of $a \simeq 0.123~{\rm fm}$. 
We build the fat link with $\alpha = 0.52$ and amplify all
link variables by a factor $u=1.28$ (in the vector term we multiply another
factor of $0.96$). This is similar (but not exactly equivalent)
to choosing the parameter $\rho$ in $D_{\rm ov-W}$; for $D_{\rm ov-HF}$
we fix $\rho =1$. These parameters are chosen as a compromise between
optimising the locality and the condition number of $A^{\dagger}A$.
Fig.\ \ref{localfig} (on top) shows that the locality is
significantly improved upon $D_{\rm ov-W}$ (at $\rho =1.6$) \cite{Stani}.

In Fig.\ \ref{localfig} (below) we illustrate the locality at strong gauge 
coupling:
at $\beta = 5.7$ ($a \simeq 0.17~{\rm fm}$) the standard overlap operator
is still barely local, if one chooses the optimal value $\rho = 1.8$.
Similarly we optimise $u =1.6$ for the HF and we still find a clear
locality at this gauge coupling --- this locality is in fact still stronger
than the one observed for $D_{\rm ov-W}$ at $\beta = 6$ (and $\rho = 1.4$,
which is locality-optimal in that case). \\
If we proceed to $\beta = 5.6$, the exponential decay evaporates 
for $D_{\rm ov-W}$, hence in this case the standard overlap formulation 
does not provide a valid Dirac operator any more. 
On the other hand, if we insert the HF at $u=1.7$ we still observe locality.
Thus we see that the overlap-HF formulation provides {\em chiral fermions
on coarser lattices.}

\section{EVALUATION OF THE LEADING LOW ENERGY CONSTANTS
IN THE $\eps$-REGIME}

In the $\eps$-regime \cite{epsreg} the correlation length exceeds 
the box length, $1/m_{\pi} > L$, and
the obser\-vables strongly depend on the topological sector \cite{LeuSmi}. 
Squeezing pions into such a tiny box may not be a realistic
situation. Still, there is a striking 
motivation for studying this regime: it allows
for an evaluation of the LEC with their values in infinite volume,
i.e.\ with their physical values. (Unfortunately, quenching
brings in logarithmic finite size effects \cite{Dam},
but there is a window of volumes with decent values nevertheless).

\begin{figure}[hbt]
  \centering
  \includegraphics[angle=270,width=1.\linewidth]{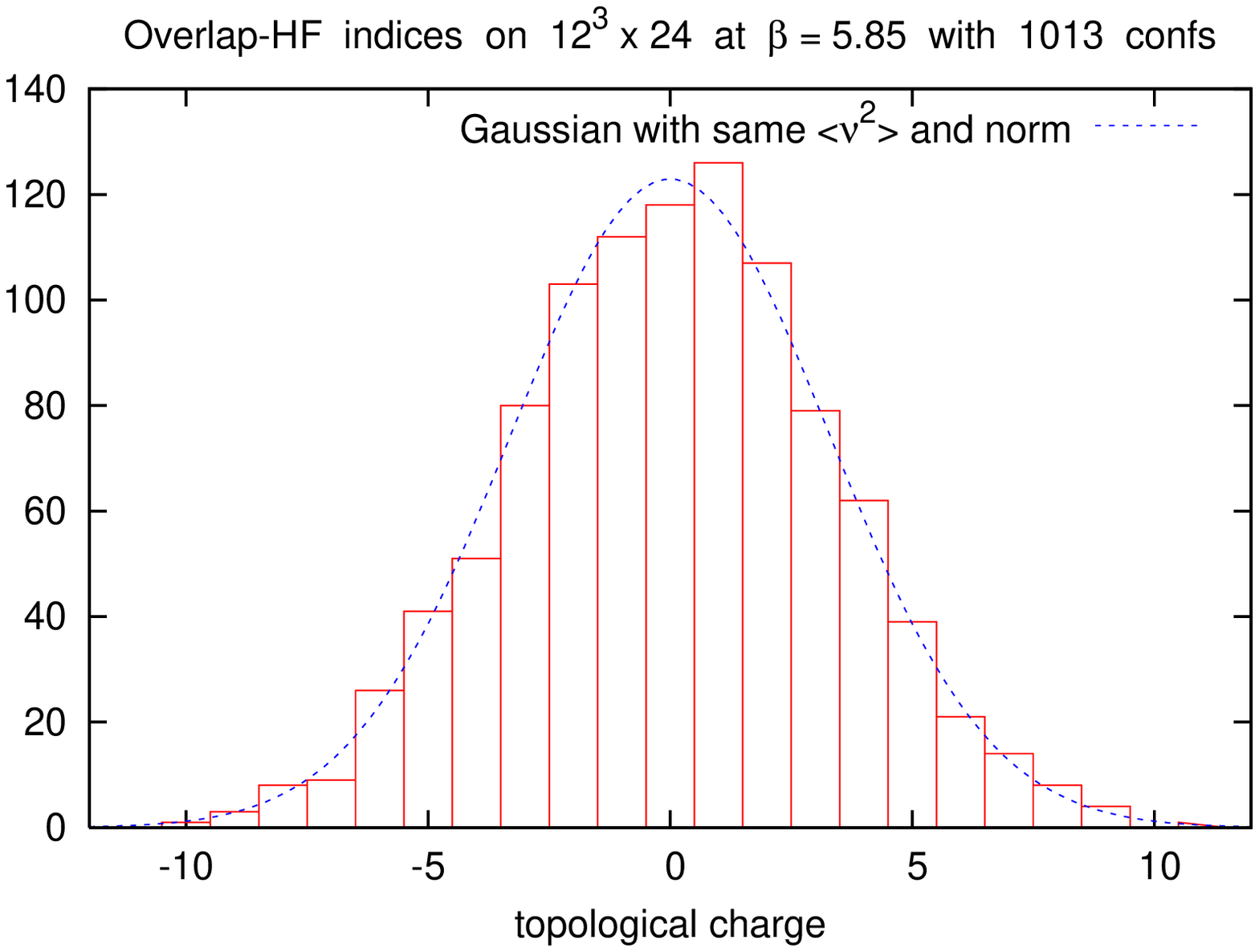}
  \includegraphics[angle=270,width=1.\linewidth]{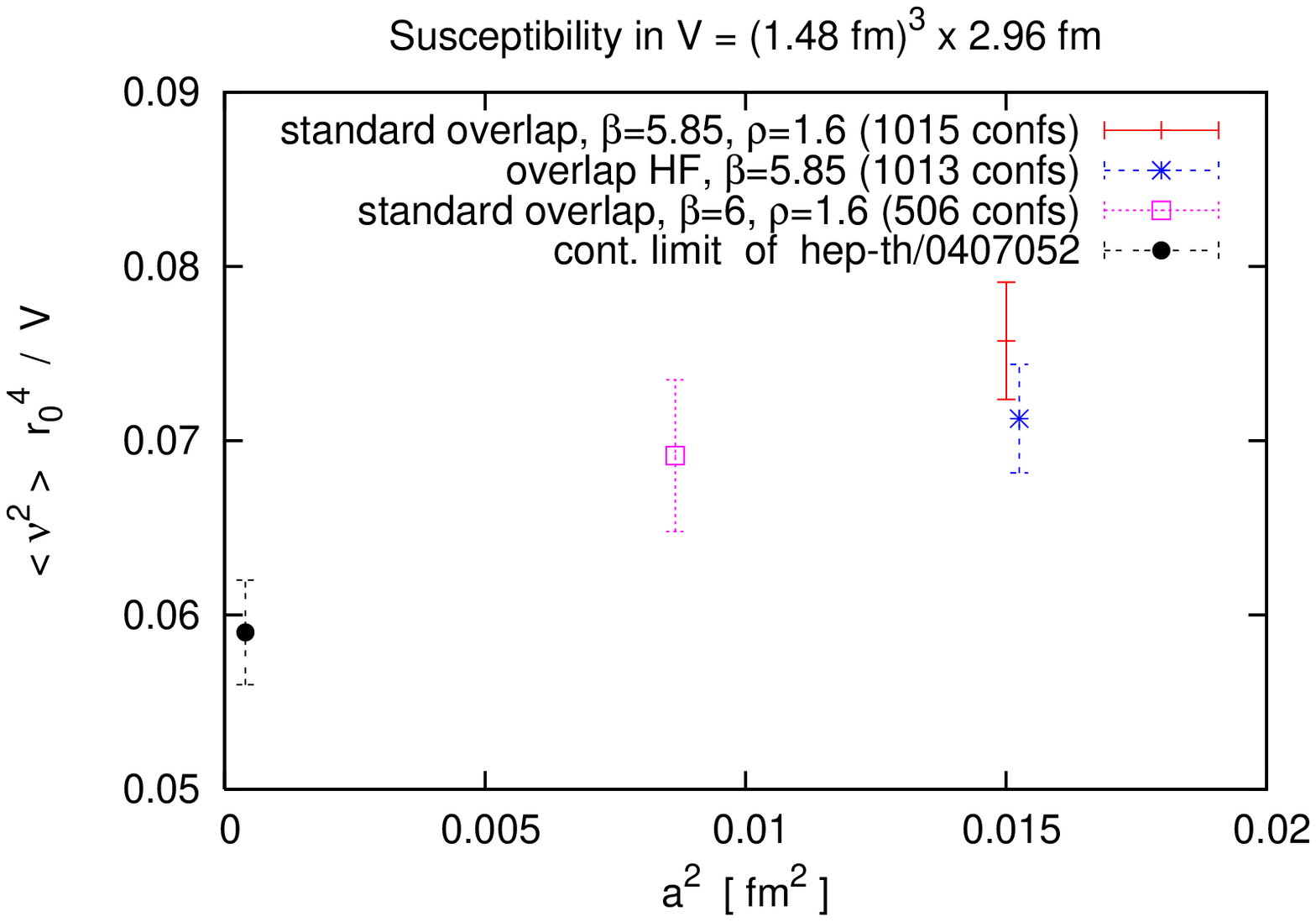}
\vspace*{-8mm}
 \caption{{\it Histogram of overlap-HF indices (on top), and
various results for the topological susceptibility (below).}}
\label{topofig}
\vspace*{-6mm}
\end{figure}

We first take a look at the overlap-HF indices: for 1013 configurations
we obtained the  histogram in Fig.\ \ref{topofig} (on top).
Fig.\ \ref{topofig} (below) shows our results for the topological 
susceptibility, compared to the continuum extrapolation of Ref.\ \cite{DGP},
which used $D_{\rm ov-W}$.  Our susceptibilities are somewhat larger,
but a rough continuum extrapolation agrees well with Ref.\ \cite{DGP}.

\vspace*{-2mm}
\subsection{Link to the chiral Random Matrix Theory}

\begin{figure}[hbt]
  \centering
\includegraphics[angle=270,width=1.\linewidth]{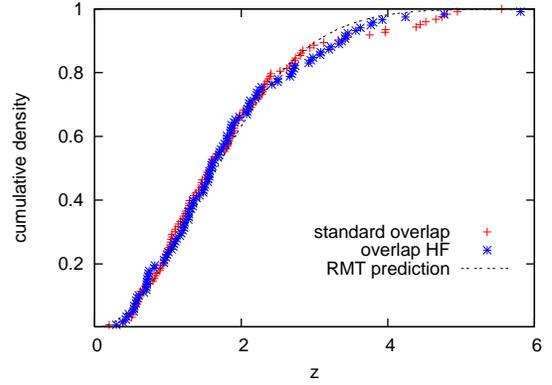}
\vspace*{-7mm}
 \caption{{\it The cumulative density of the first Dirac eigenvalue
$\lambda_{1}$ in the sector $\nu =0$ --- 
stereographically projected and re-scaled 
with $\Sigma V$ at optimal $\Sigma$ --- 
compared to the RMT prediction.}}
\label{RMTfig}
\vspace*{-5mm}
\end{figure}

Chiral Random Matrix Theory (RMT) conjectures the densities 
$\rho_{n}^{(\nu )}(z)$ of the lowest
Dirac eigenvalues $\lambda$ in the $\eps$-regime \cite{DamNish},
where $z := \lambda \Sigma V$. $n= 1,2, \dots$ numerates the lowest 
non-zero eigenvalues and $\nu$ is the fermion index, 
which is identified with the topological charge \cite{Has-Lusch}.
This RMT conjecture holds to a good precision
for the lowest $n$ and $| \nu |$,
if $L$ exceeds some lower limit (about $1.1 \dots 1.5$ fm,
depending on the exact criterion) \cite{RMT}. 
Then the fit determines the scalar condensate $\Sigma$.
From the cumulative
density of the first eigenvalue in the topologically neutral sector,
the fit implies a chiral condensate of $\Sigma = (299 (8)~{\rm MeV})^{3}$
for the overlap HF, in good agreement with the result of
$\Sigma = (300(7)~{\rm MeV})^{3}$ for the standard overlap operator,
see Fig.\ \ref{RMTfig}.

\vspace*{-2mm}
\subsection{Zero-mode contributions to the pseudoscalar correlation}

We stay in the chiral limit and evaluate
the time derivative of the zero-mode contribution 
to the correlation of the pseudo-scalar density $P = \bar \psi \gamma_5 \psi$,
\begin{eqnarray}
\hspace*{-5mm} &&
C'_{|\nu|}(t) = \frac{d}{dt}\sum_{\vec{x},i,j}\Big\langle \langle
j; \vec{x}, t|i; \vec{x},t \rangle \langle i; 0|j; 0\rangle
\Big\rangle_{\nu} \ , \nonumber \\ 
\hspace*{-5mm} &&
{\rm with} \qquad D_{\rm ov}^{(0)} \, |i; \vec{x},t\rangle = 0 \ .
\end{eqnarray}
This quantity was computed analytically to next-to-next-to-leading order
(NNLO) in 
quenched $\chi$PT in Ref.~\cite{zeromode}, which also
presented a numerical study with $D_{\rm ov-W}$. 
We extend this study by including the overlap-HF.

To NNLO the quenched chiral Lagrangian reads
\begin{eqnarray*}
\hspace*{-7mm} &&
{\cal L^{\rm NNLO}} = \frac{F^2_{\pi}}{4}{\rm Str}(\partial_\mu U
\partial_\mu U^\dagger ) -\frac{\Sigma}{2}{\rm
Str}(MU + {\rm h.c.} ) \\ 
\hspace*{-7mm} && - iK\Phi_0{\rm Str}(MU -{\rm h.c.})
+\frac{m^2_0}{2N_c}\Phi^2_0
+ \frac{\alpha_0}{2N_c}(\partial_\mu
\Phi_0)^2\, ,
\end{eqnarray*}
where $U \in SU(N_f|N_f)\otimes SU(N_f|N_f)/SU(N_f|N_f)$, 
``Str'' is the corresponding ``supertrace'', 
and $M$ is the diagonal matrix of fermion masses.
The peculiarity of the quenched chiral
Lagrangian is the presence of a scalar field $\Phi_0$.
${\cal L^{\rm NNLO}}$ contains five quenched LEC: 
$F_\pi$, $\Sigma$, $K$, $m_0$ and $\alpha_0$. 

We consider the first term in the Taylor expansion of
$C'_{|\nu|}(t)$ around $T/2$. In a box $L^3\times T$ it takes
the form~\cite{zeromode}
\begin{equation}
\frac{C'_{|\nu|}(t)}{L^2} = D^{\rm NNLO}_{|\nu|}(F_\pi, \langle
\nu^2\rangle,\alpha)\frac{s}{T} +{\cal O}\left ( \left
(\frac{s}{T}\right )^3\right )
\label{Taylor}
\end{equation}
with \ $ s:=t-T/2$ and $\alpha := \alpha_0
-2N_{c}KF_\pi/\Sigma \,$. The dependence on $\langle \nu^2\rangle$
originates from the Witten-Veneziano formula~\cite{WV}, and 
$m_{0}$ was counted in ${\cal O}(\varepsilon )$.

We analysed the zero-modes that we collected 
at $\beta=5.85$ on a $12^3\times 24$ lattice, and at
$\beta=6$ on $16^3\times 32$. In
both cases, the physical volume amounts again to
$(1.48\,{\rm fm})^3\times 2.96 \,{\rm fm}$.  
This is expected to be sufficiently large,
referring to our
experience with the chiral Random Matrix Theory~\cite{RMT} and to
the behaviour of the axial correlators in the
$\epsilon$-regime (cf.\ Subsection 3.3). 
Our statistics is listed in Table~\ref{tab1}.
\begin{table*}[htb]
\begin{center}
\begin{tabular}{|l|c|c||c|c||c|c|}
\hline
 & $\rho$ & $\langle \nu^2 \rangle$ &
 $|\nu|=1$ & $|\nu|=2$ & $F_\pi$ [MeV] & $\alpha$ \\
 \hline
overlap-HF $12^3\times 24$, $\beta=5.85$ & 1 & $10.8$ &
$221$ & $192$ & $80 \pm 14$ & $-17 \pm 10$ \\
 \hline
standard ov. $12^3\times 24$, $\beta=5.85$ & 1.6& $11.5$ &
$232$ & $180$ & $74 \pm 10$ & $-19 \pm 8 ~$ \\
 \hline
standard ov. $16^3\times 32$, $\beta=6$ & 1.6& $10.5$ &
$115$ & $94$ & $75 \pm 24$ & $-21 \pm 15$ \\
 \hline
\end{tabular}
\end{center}
\vspace*{-2mm}
\caption{{\it Statistics and results for $F_\pi$ and $\alpha$,
based on the zero-mode contributions to $\langle PP \rangle$.
The values for $F_{\pi}$ and $\alpha$ in the last two columns
correspond to the fitting range $s_{\rm max}=1$.}} 
\label{tab1}
\vspace*{-2mm}
\end{table*}
We fixed the value of $\langle \nu^2 \rangle$ from our simulations,
since it can deviate significantly from the continuum extra\-polation,
as Fig.\ \ref{topofig} (below) shows.

\begin{figure}[hbt]
\vspace*{-13mm}
\begin{center}
\includegraphics[width=0.33\textwidth,angle=-90]{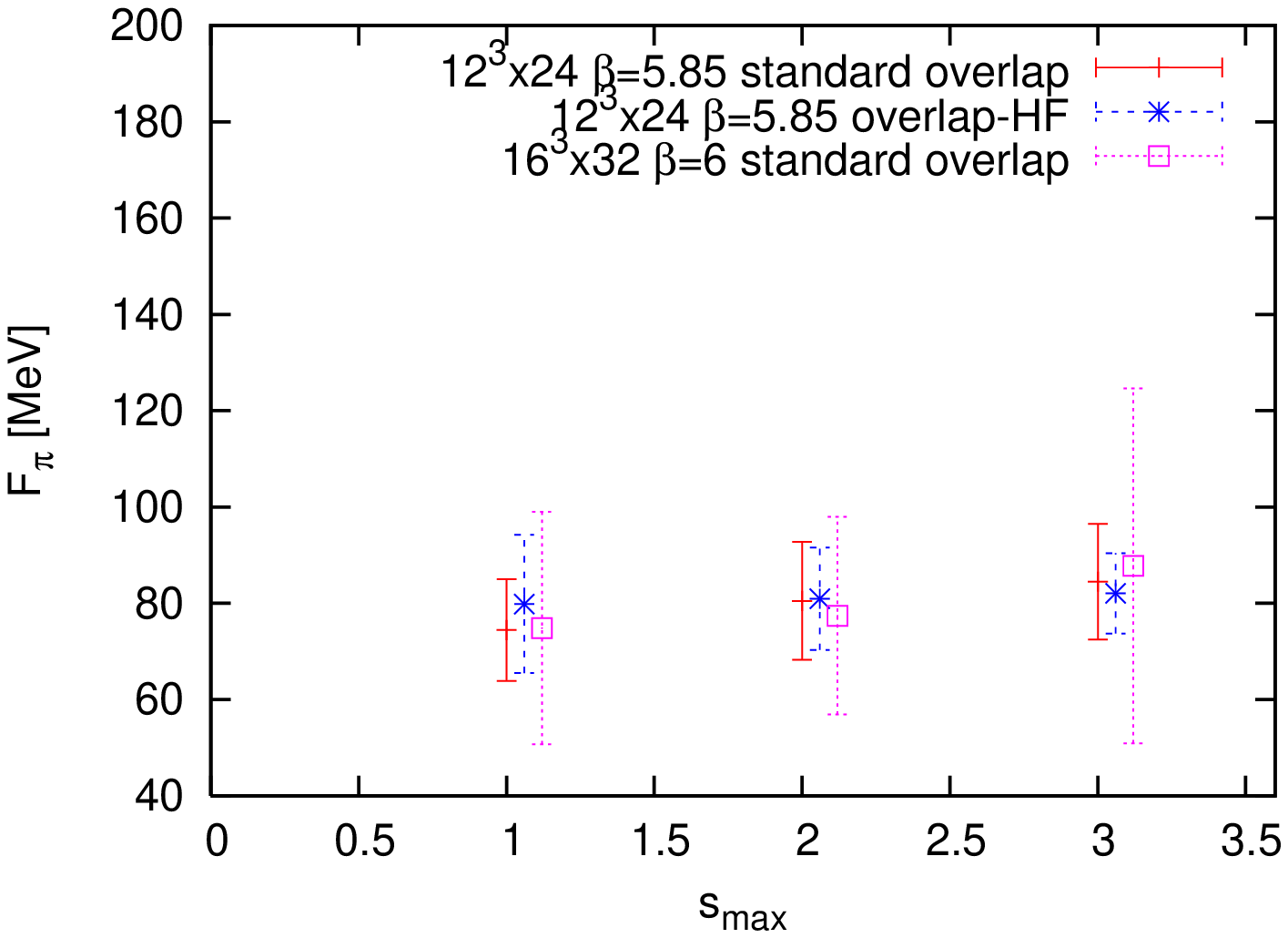}
\includegraphics[width=0.33\textwidth,angle=-90]{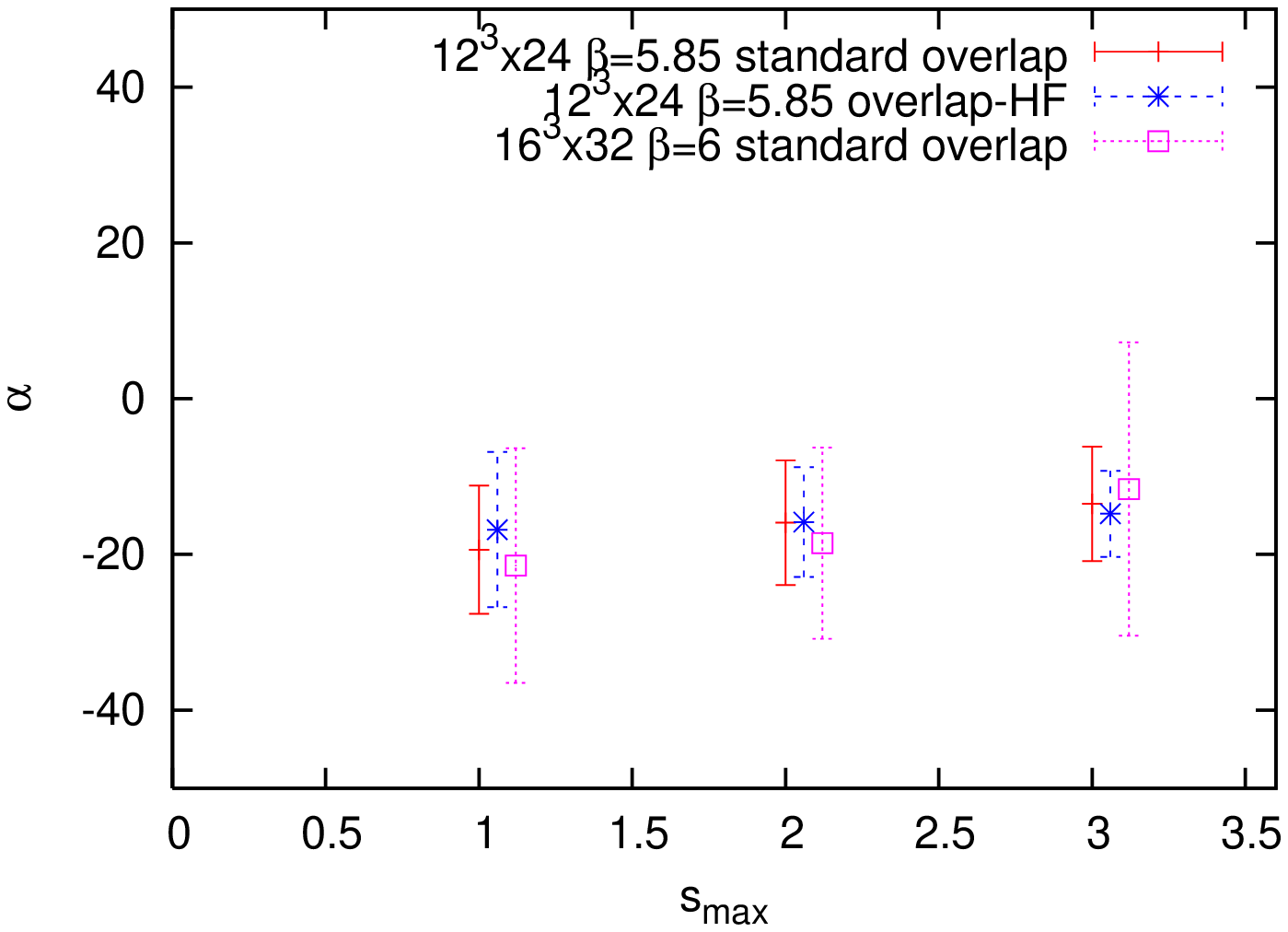}
\end{center}
\vspace*{-9mm}
\caption{{\it The Low Energy Constants $F_\pi$ and $\alpha$,
determined from the zero-mode contributions to $\langle PP \rangle$,
against the width of the fitting range $s_{\rm max}$.}} 
\label{fig1}
\vspace*{-8mm}
\end{figure}

Then we performed a combined two parameter fit of the leading term in
eq.~(\ref{Taylor}) to our data in the
topological sectors $|\nu|=1,2$ over the intervals $s \in
[-s_{\rm max} ,s_{\rm max}]$, with
$s_{\rm max}=1 \dots 3$. Our results are shown
in Fig.~\ref{fig1} and in Table~\ref{tab1}.
Our values for $F_{\pi}$ and $\alpha$ are a little lower than 
the results reported in Ref.\ \cite{zeromode}
(for the Neuberger operator in isotropic boxes).

However, we see that our values of
$F_\pi$ and $\alpha$ 
agree well
for different overlap fermions and for two lattice spacings.
This result for $F_{\pi}$ is close to the physical value
--- extrapolated to the chiral limit --- of $86 ~{\rm MeV}$ \cite{CD},
but below typical quenched values, see Subsection 3.3.

\vspace*{-0mm}

\subsection{Axial vector correlations}

\begin{figure}[hbt]
  \centering
  \includegraphics[angle=270,width=1.\linewidth]{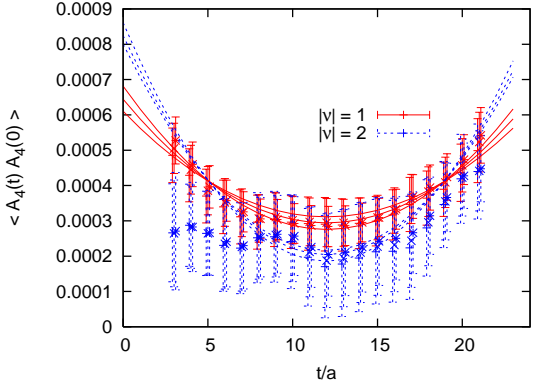}
\vspace*{-7mm}
\caption{{\it Results for the axial vector correlator
at $am_{q} = 0.001$, $0.003$ and $0.005$
in the sectors $| \nu | =1$ (red, larger around the centre)
and $2$ (blue), and fits to the formulae 
of quenched $\chi$PT.
(Our quark masses correspond to $\Sigma V m_{q} = 0.83,$ $2.5$ 
and $4.2$.)}}
\label{AAfig}
\vspace*{-7mm}
\end{figure}

In quenched $\chi$PT, the axial vector correlator 
depends to leading order only on the LEC
$\Sigma $ and $F_{\pi}$ \cite{qXPT2}. 
The prediction for $\la A_{4} (t) A_{4} (0)\ra$
(with $A_{4}(t):= \sum_{\vec x} \bar \psi(\vec x, t) \gamma_{5} 
\gamma_{4} \psi (\vec x , t)$, i.e.\ the bare axial current
at $\vec p = \vec 0$) 
is a parabola with a minimum at $t = T/2$, 
where $F_{\pi}^{2} /T$ enters as an additive constant. In a previous study
we observed that $L$ should again be above $1 ~{\rm fm}$ for this
prediction to work, and that the
history in the sector $\nu =0$ may be plagued by spikes,
related to the occurrence of very small non-zero eigenvalues \cite{AA}.
(This leads to the requirement of a large statistics, but
a method called Low Mode Averaging was then invented as an attempt
to overcome this problem \cite{LMA}.)
In the non-trivial sectors, $\Sigma$ can hardly be determined,
but $F_{\pi}$ can be evaluated \cite{prelim,Jap}. 
Fig.\ \ref{AAfig} shows our
current results with three (bare) quark masses
$m_{q}$ in the $\eps$-regime, in the sector $| \nu | = 1$ and $2$. 
The (flavour degenerated) mass is incorporated in the overlap operator as
\vspace{-1mm}
\beq
\vspace{-2mm}
D_{\rm ov}(m_{q}) = \Big( 1 - \frac{m_{q}}{2 \rho} \Big) 
D_{\rm ov}^{(0)} + m_{q} \ .
\vspace{-2mm}
\eeq
\vspace{-1mm}
Our result is based on
45 propagators for each mass in the sector $\vert \nu \vert =1$,
and 10  propagators for each mass in the sector $\vert \nu \vert =2$.
Inserting the RMT result of $\Sigma \simeq (299~{\rm MeV})^{3}$, 
as well as the renormalisation constant $Z_{A}=1.17$ (see Section 4),
a global fit leads to ${\bf{F_{\pi} = (113 \pm 7)}} ~ {\rm{\bf{ MeV}}}$, 
in agreement with other quenched results \cite{LMA,zeromode,Jap}.

\vspace*{-2mm}
\section{RESULTS IN THE $p$-REGIME}

At last we present our results in the $p$-regime, which is characterised 
by a box length $L~\gg~1/m_{\pi}$, so that the $p$-expansion of $\chi$PT
\cite{preg} applies.
We consider once more $\beta=5.85$, a lattice of size $12^{3} \times 24$
and the bare quark masses $a m_{q} = 0.01, \ 0.02, \ 0.04,\ 0.06, \
0.08$ and $0.1$ (in physical units: $m_{q}
= 16.1 ~ {\rm MeV} \dots 161~{\rm MeV}$). 
We computed 100 overlap-HF propagators, and we first evaluated 
$m_{\pi}$ in three different ways:\\
(1) From the pseudoscalar correlator $\la P(t)P(0) \ra$. \\
(2) From the axial vector correlator
$\la A_{4}(t) A_{4}(0) \ra$. \\
(3) From $\la P(t)P(0) - S(t)S(0) \ra$, where 
$S = \bar \psi \psi$. The subtraction of the scalar density is
useful at small $m_{q}$ to avoid the contamination by zero modes.

\begin{figure}[hbt]
  \centering
  \includegraphics[angle=270,width=1.\linewidth]{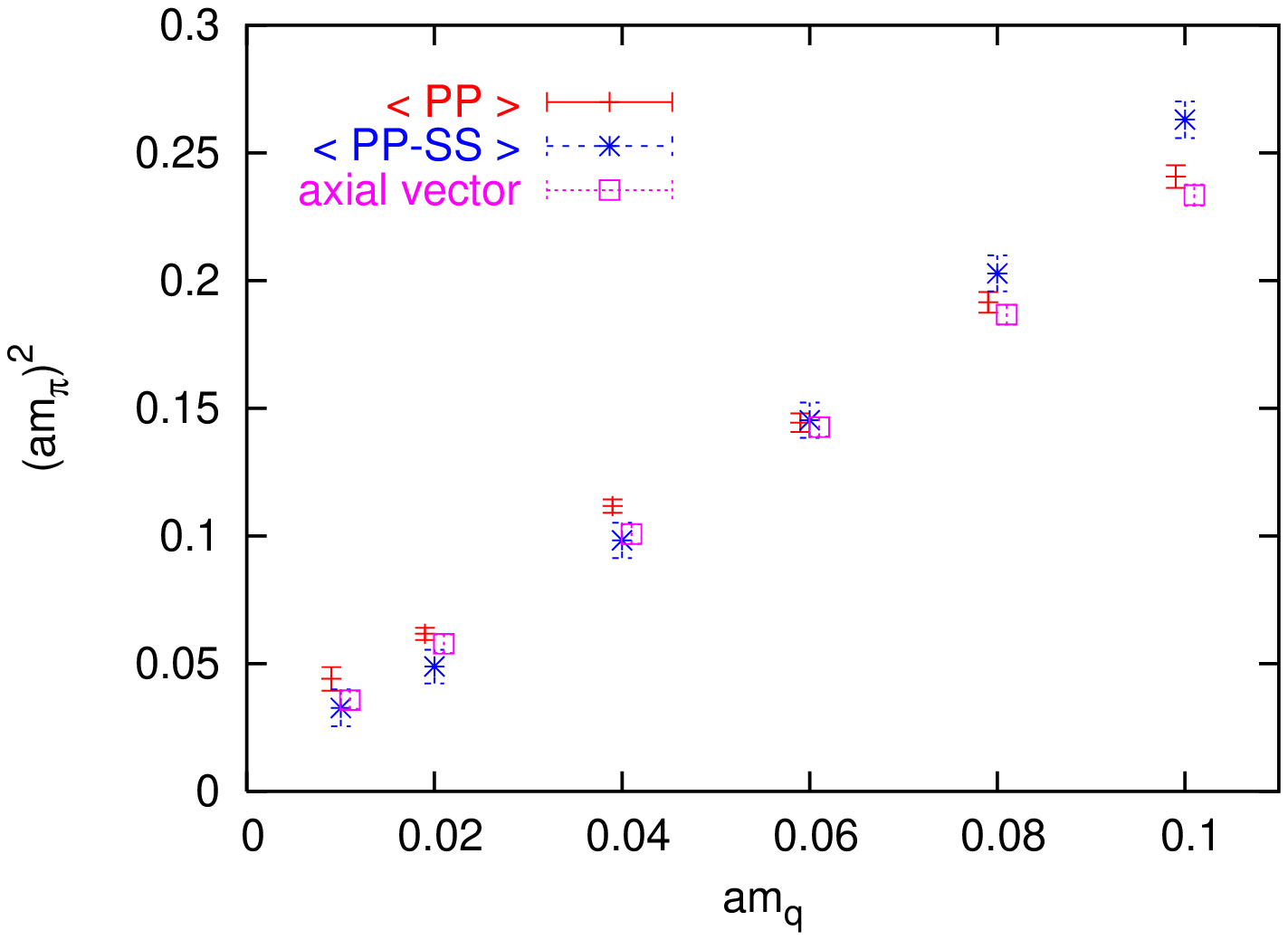}
  \includegraphics[angle=270,width=1.\linewidth]{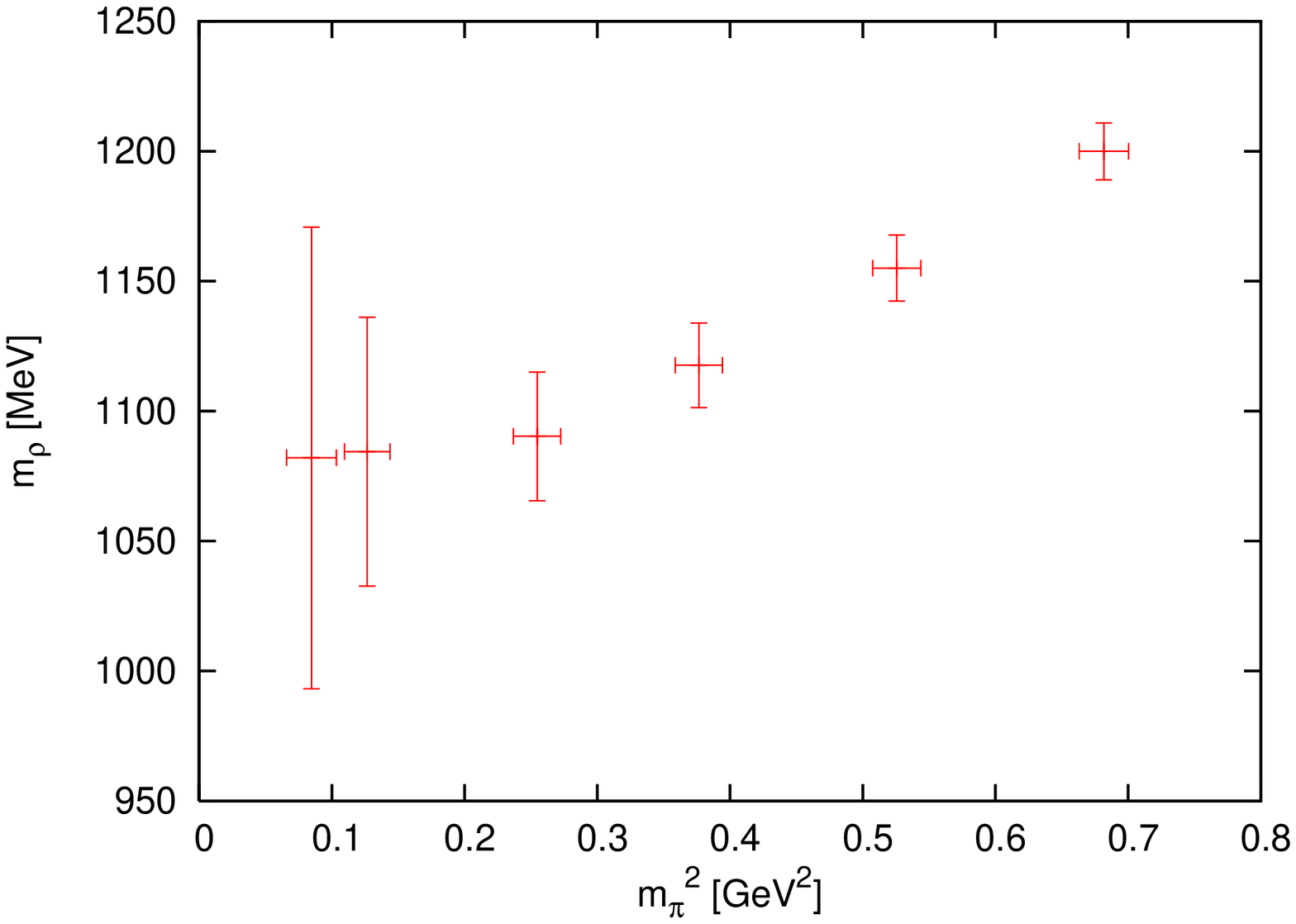}
\vspace*{-7mm}
 \caption{{\it On top: The pion mass evaluated in various ways.
Below: the $\rho$-meson mass.}}
\label{pirhofig}
\vspace*{-6mm}
\end{figure}

The results are shown in Fig.\ \ref{pirhofig} (on top):
they follow well the expected behaviour
$a m_{\pi}^{2} \propto m_{q}$, in particular for
$\la PP - SS \ra$.
Our smallest pion mass in this plot, $m_{\pi} \simeq 289 ~{\rm MeV}$,
has Compton wave length $\approx L/2$, hence
this point is at the edge of the $p$-regime.

Fig.\ \ref{pirhofig} (below) shows the corresponding results 
for the vector meson mass, with a (linear) chiral extrapolation to
$m_{\rho} = 1017(39) ~{\rm MeV}$.
Generally, quenched results tend to be above
the physical $\rho$-mass. Our result is in agreement with a
similar study using $D_{\rm ov-W}$ \cite{XLF},
but somewhat above the (quenched) world data for
$m_{\rho}$ \cite{Arifa} (which are mostly obtained with
chiral symmetry breaking lattice fermions).

\begin{figure}[hbt]
  \centering
  \includegraphics[angle=270,width=1.\linewidth]{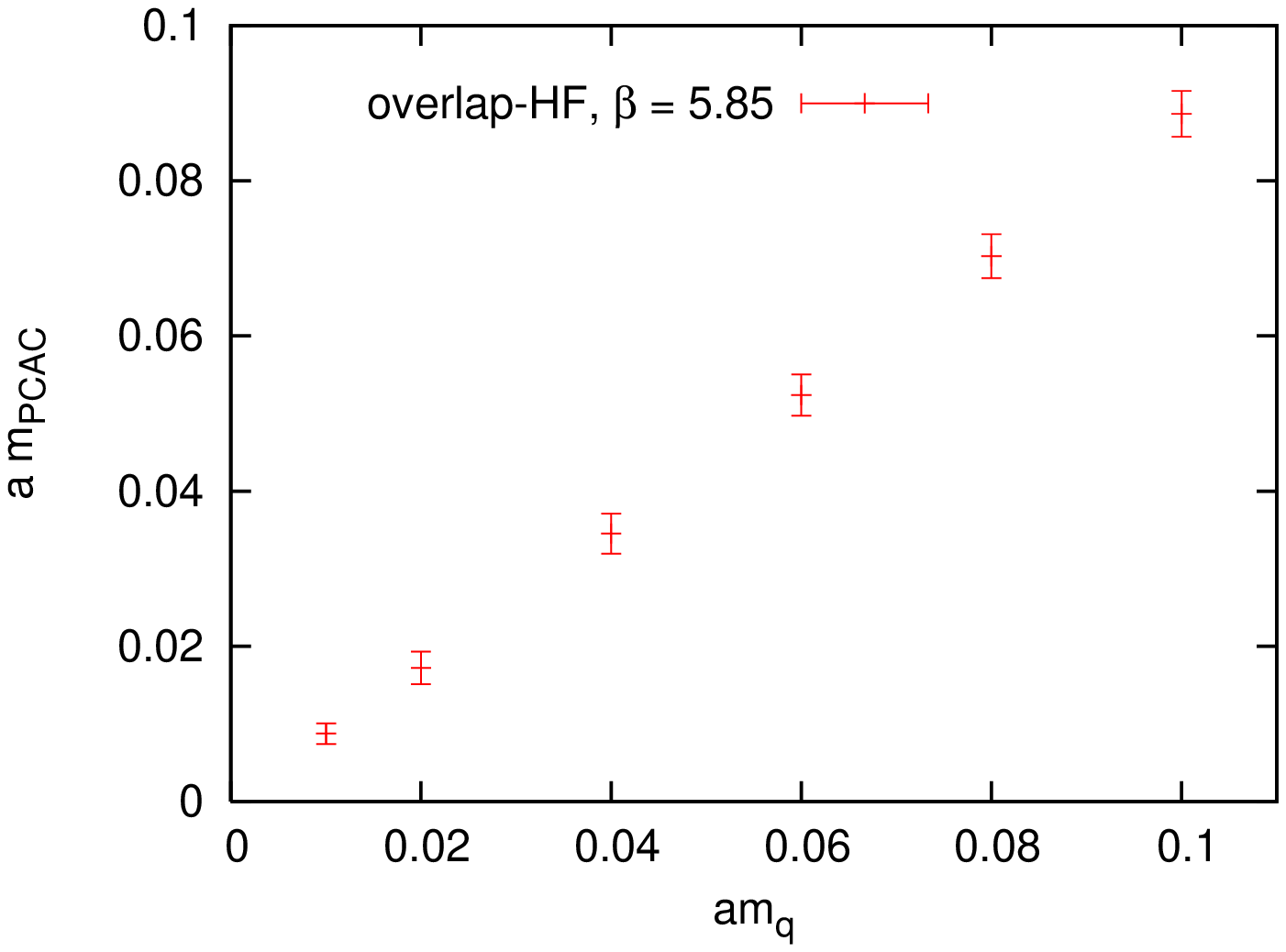}
  \includegraphics[angle=270,width=1.\linewidth]{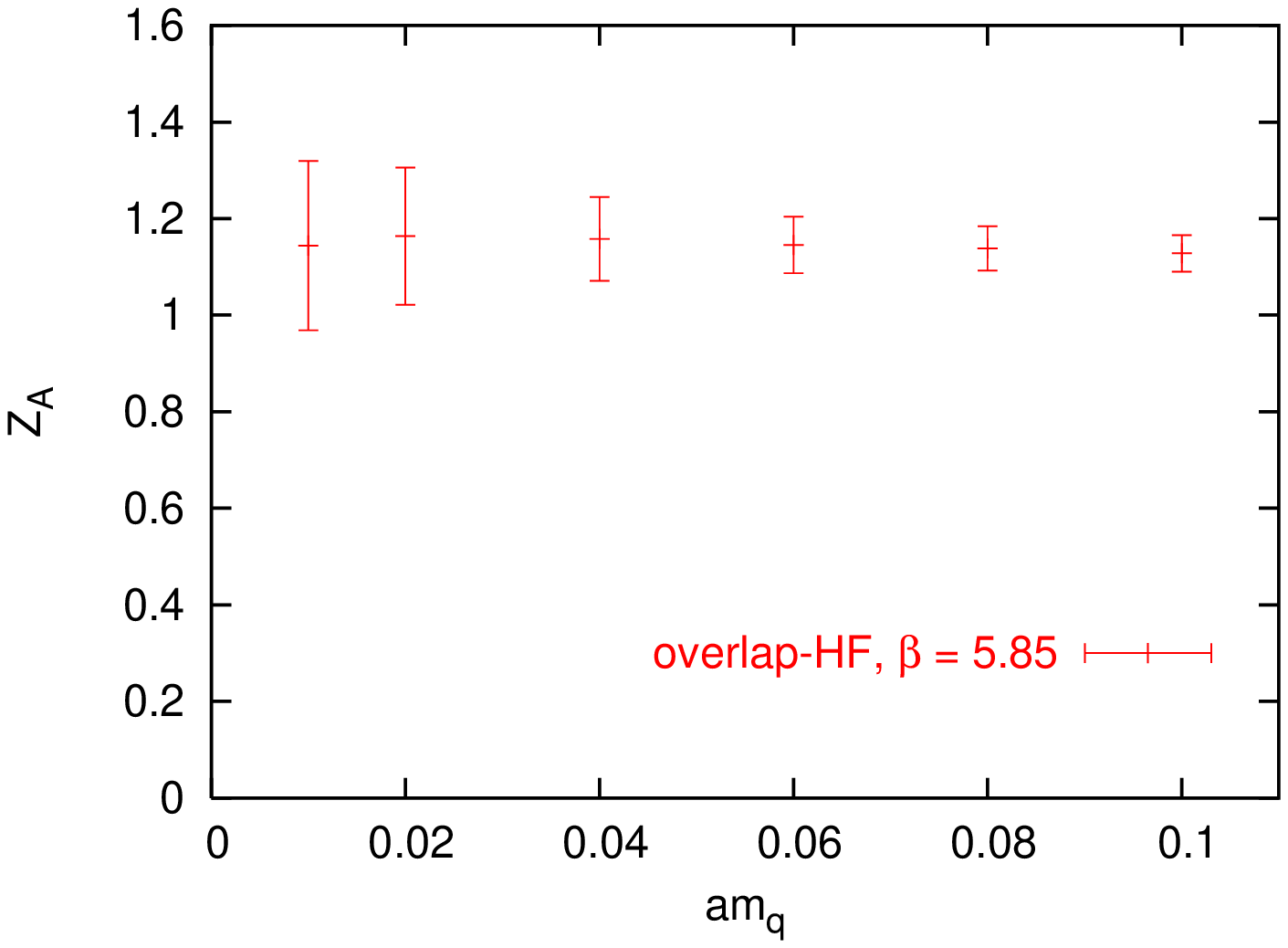}
\vspace*{-7mm}
 \caption{{\it The PCAC quark mass of eq.\ (\ref{AWI}), and
the renormalisation constant $Z_{A} = m_{q}/ m_{\rm PCAC}$.}}
\label{mPCAC_ZAfig}
\vspace*{-7mm}
\end{figure}
In Fig.\ \ref{mPCAC_ZAfig} (on top)
we show the PCAC quark mass obtained from the 
axial Ward identity,
\beq  \label{AWI}
m_{\rm PCAC} = \frac{\sum_{\vec x} \la \partial_{t} A^{\dagger}_{4}(x)
P(0) \ra }{2 \, [ \, \sum_{\vec x} \la P^{\dagger} (x) P(0) \ra \, ]} \ .
\eeq
\begin{figure}[hbt]
  \centering
\includegraphics[angle=270,width=1.\linewidth]{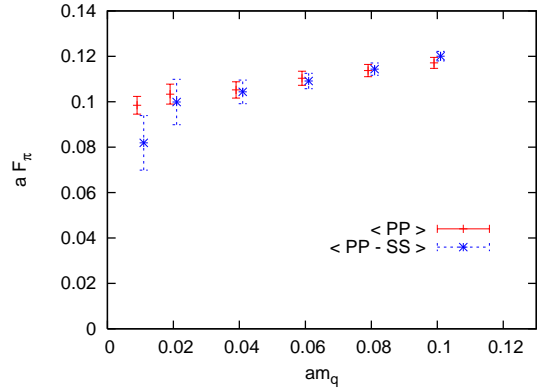}
\vspace*{-8mm}
\caption{{\it The pion decay constant from a direct evaluation in the
$p$-regime, using the overlap-HF.}}
\label{Fpifig}
\vspace*{-7mm}
\end{figure}
\noindent
We observe also here a behaviour which is
nearly linear in $m_{q}$, with a chiral extrapolation to \\
$am_{\rm PCAC}(m_{q}=0) = - 0.00029(64)$.\\
Remarkably, $m_{\rm PCAC}$ is close to $m_{q}$, which is {\em not} 
the case for $D_{\rm ov-W}$ \cite{XLF}. 
Consequently the renormalisation
constant $Z_{A} = m_{q}/ m_{\rm PCAC}$ is close to 1; it
has the chiral extrapolation $Z_{A} = 1.17(2)$, which has been
used already in Subsection 3.3.
This is in contrast to the large $Z_{A}$ factors
obtained for $D_{\rm ov-W}$: at $\beta =5.85$
($\rho =1.6$), $Z_{A} \simeq 1.45$ was measured \cite{XLF,Jap}, 
and at $\beta =6$ ($\rho =1.4$) even $Z_{A} \simeq 1.55$ \cite{Berruto}.

Finally we reconsider the pion decay constant, now evaluated in the
direct way, according to 
\beq
F_{\pi} = 
\frac{2 m_{q}}{m_{\pi}^{2}} \vert \la 0 \vert P \vert \pi \ra \vert \ ,
\eeq
by using either $PP$ or $PP-SS$, see Fig.\ \ref{Fpifig}.
The linear extrapolation to $m_{q}=0$ yields
${\bf{ F_{\pi , PP} = 111.5 (2.5)}} {\rm{\bf{~MeV}}}$, resp.
${\bf{ F_{\pi , PP-SS} = 104(9) }}~{\rm{\bf{ MeV}}}$, which is
compatible with the result in Subsection 3.2.
In particular the behaviour of the 
$\la PP-SS \ra$ result for $F_{\pi}$ at small $m_q$ calls for
an evaluation at yet smaller quark masses, which was reported 
in Section 3.

\section{CONCLUSIONS}

The HF operator is approximately chiral and it has favourable
properties in particular in thermodynamics. It can be inserted
into the overlap formula, which yields the overlap-HF. 
This is a formulation of chiral lattice fermions
with better rotation symmetry and locality than the standard formulation.
Therefore it provides access to chiral
fermions on coarser lattices. \\

In the $\epsilon$-regime we evaluated ---
at tiny quark masses --- the pion decay constant from the axial
vector correlator. Directly in the chiral limit we obtained
results for the chiral condensate from the Dirac spectrum.
Moreover, from the zero-mode contribution to the pseudoscalar correlator,
we found again values of $F_\pi$ and the parameter $\alpha$
(a quenching specific LEC).
The results are very similar for two lattice spacings and
different overlap operators. 
Our result for $F_{\pi}$ obtained with this method 
(in the given volume) is close to the
physical value (in the chiral limit), but below other quenched values.
\footnote{The two methods applied for the axial vector and the pseudoscalar
correlator differ by a subtlety in the quenched counting rules for the
$\epsilon$-expansion.} \\
We add that a topology conserving gauge action could be helpful in that regime
\cite{topogauge}. \\

In the $p$-regime we obtained results for $m_{\pi}$, $m_{\rho}$ and 
$F_{\pi}$. Compared to the standard overlap fermion, $m_{\rm PCAC}$ is closer to
the bare quark mass $m_{q}$, hence $Z_{A}$ is much closer to $1$,
which is pleasant for a connection to perturbation theory.
Here we measured $F_{\pi}$ directly, and the chiral extrapolation of
this result agrees (within the errors) with the
value obtained in the $\epsilon$-regime from the axial vector correlator.

\newpage

\noindent
{\bf Acknowledgements} ~ W.B. would like to thank the organisers
of this Workshop on Computational Hadron Physics for their kind
hospitality in Cyprus. 
We both thank A.\ Ali Khan, W.\ Cas\-sing, C.\ Fischer and
P.\ Watson for helpful comments on the $\rho$-mass, and
S.\ D\"{u}rr, H.\ Fukaya, P. Hasenfratz, K.-I.\ Nagai, M.\ Papinutto 
and C.\ Urbach for inspiring communications.

\end{document}